\documentclass[aps,showpacs,twocolumn,floats,superscriptaddress,floatfix]{revtex4}
\usepackage{graphicx,bm,times,url}
\usepackage{url}
\graphicspath{{./fig/}{./png/}}
\newcommand{\BoldVec}[1]{\mathchoice%
  {\mbox{\boldmath $\displaystyle     #1$}}%
  {\mbox{\boldmath $\textstyle        #1$}}%
  {\mbox{\boldmath $\scriptstyle      #1$}}%
  {\mbox{\boldmath $\scriptscriptstyle#1$}}%
}
\newcommand{\EQ}{\begin{equation}}
\newcommand{\EN}{\end{equation}}
\newcommand{\EQA}{\begin{eqnarray}}
\newcommand{\ENA}{\end{eqnarray}}

\newcommand{\Eq}[1]{Eq.~(\ref{#1})}
\newcommand{\Eqs}[2]{Eqs.~(\ref{#1}) and~(\ref{#2})}

\newcommand{\Sec}[1]{Sec.~\ref{#1}}

\newcommand{\Fig}[1]{Fig.~\ref{#1}}

\newcommand{\bra}[1]{\langle #1\rangle}

\newcommand{\meanC}{\overline{C}}
\newcommand{\meanU}{\overline{U}}
\newcommand{\meanW}{\overline{W}}
\newcommand{\meanF}{\overline{\cal F}}

\newcommand{\meanUU}{\overline{\bm{U}}}
\newcommand{\meanWW}{\overline{\bm{W}}}

\newcommand{\meanAAAA}{\overline{\mbox{\boldmath ${\mathsf A}$}} {}}
\newcommand{\meanSSSS}{\overline{\mbox{\boldmath ${\mathsf S}$}} {}}
\newcommand{\meanAAA}{\overline{\mathsf{A}}}
\newcommand{\meanSSS}{\overline{\mathsf{S}}}

\newcommand{\meanEMF}{\overline{\mbox{\boldmath ${\cal E}$}} {}}
{}
{}
\newcommand{\meanFFFF}{\overline{\mbox{\boldmath ${\cal F}$}}{}}{}
\newcommand{\meanFFF}{\overline{\cal F}}
\newcommand{\hatOO}{\hat{\bm{\Omega}}}
\newcommand{\Oh}{\hat{\Omega}}

\newcommand{\yyy}{\hat{\mbox{\boldmath $y$}} {}}

\newcommand{\uu}{\BoldVec{u} {}}

\newcommand{\UU}{\BoldVec{U} {}}

\newcommand{\ff}{\BoldVec{f} {}}

\newcommand{\FF}{\BoldVec{F} {}}

\newcommand{\kk}{\BoldVec{k} {}}

\newcommand{\nab}{\BoldVec{\nabla} {}}
\newcommand{\OO}{\BoldVec{\Omega} {}}

\newcommand{\RRRR}{\bm{\mathsf{R}}}
\newcommand{\SSSS}{\bm{\mathsf{S}}}
\newcommand{\LLLL}{\mbox{\boldmath ${\sf L}$} {}}

\newcommand{\ii}{{\rm i}}

\newcommand{\diag}{{\rm diag}  \, {}}
\newcommand{\DD}{{\rm D} {}}
\newcommand{\dd}{{\rm d} {}}
\newcommand{\const}{{\rm const}  {}}

\def\Ma{\mbox{\rm Ma}}
\def\Sh{\mbox{\rm Sh}}

\def\St{\mbox{\rm St}}
\def\Rey{\mbox{\rm Re}}
\def\Pe{\mbox{\rm Pe}}
\def\Sc{\mbox{\rm Sc}}
\def\cs{c_{\rm s}}
\def\kf{k_{\rm f}}
\def\urms{u_{\rm rms}}

\def\kappaOO{\kappa_{\Omega\Omega}}
\def\kappaO{\kappa_{\Omega}}
\def\kappat{\kappa_{\rm t}}
\def\kappatz{\kappa_{\rm t0}}

\def\half{{\textstyle{1\over2}}}

\def\onethird{{\textstyle{1\over3}}}

\def\quarter{{\textstyle{1\over4}}}
\newcommand{\yan}[3]{, Astron. Nachr. {\bf #2}, #3 (#1).}

\newcommand{\yana}[3]{, Astron. Astrophys. {\bf #2}, #3 (#1).}

\newcommand{\yjetp}[3]{, Sov. Phys. JETP {\bf #2}, #3 (#1).}

\newcommand{\ymn}[3]{, Mon.\ Not.\ R.\ Astron.\ Soc.\ {\bf #2}, #3 (#1).}

\newcommand{\ynat}[3]{, Nature {\bf #2}, #3 (#1).}
\newcommand{\yjfm}[3]{, J. Fluid Mech. {\bf #2}, #3 (#1).}

\newcommand{\yrmp}[3]{, Rev.\ Mod.\ Phys.\ {\bf #2}, #3 (#1).}

\newcommand{\ypre}[3]{, Phys.\ Rev.\ E {\bf #2}, #3 (#1).}

\newcommand{\yaj}[3]{, Astronom. J. {\bf #2}, #3 (#1).}
\newcommand{\yapj}[3]{, Astrophys. J. {\bf #2}, #3 (#1).}

\newcommand{\yapjl}[3]{, Astrophys. J. {\bf #2}, #3 (#1).}
\newcommand{\ypp}[3]{, Phys. Plasmas {\bf #2}, #3 (#1).}

\newcommand{\ypf}[3]{, Phys. Fluids {\bf #2}, #3 (#1).}

\newcommand{\ygafd}[3]{, Geophys. Astrophys. Fluid Dynam. {\bf #2}, #3 (#1).}

\newcommand{\yjour}[4]{, #2 {\bf #3}, #4 (#1).}

\newcommand{\ybook}[3]{, {\em #2}. #3 (#1).}

\begin{document}
\preprint{NORDITA 2009-37}

\title{Calibrating passive scalar transport in shear-flow turbulence}

\author{Enik\H{o} J. M. Madarassy}
\affiliation{NORDITA, AlbaNova University Center, Roslagstullsbacken 23,
SE-10691 Stockholm, Sweden}

\author{Axel Brandenburg}
\affiliation{NORDITA, AlbaNova University Center, Roslagstullsbacken 23,
SE-10691 Stockholm, Sweden}
\affiliation{Department of Astronomy, AlbaNova University Center,
Stockholm University, SE 10691 Stockholm, Sweden}

\date{\today,~ $ $Revision: 1.32 $ $}
\begin{abstract}
The turbulent diffusivity tensor is determined for linear shear flow
turbulence using numerical simulations.
For moderately strong shear, the diagonal components are found to
increase quadratically with Peclet and Reynolds numbers
below about 10 and then become constant.
The diffusivity tensor is found to have components proportional to the
symmetric and antisymmetric parts of the velocity gradient matrix,
as well as products of these.
All components decrease with the wave number of the mean field in a
Lorentzian fashion.
The components of the diffusivity tensor are found not to depend
significantly on the presence of helicity in the turbulence.
The signs of the leading terms in the expression for the diffusion
tensor are found to be in good agreement with estimates based on a simple
closure assumption.
\end{abstract}

\maketitle

\section{Introduction}

In a turbulent flow, chemicals tend to be mixed more effectively
than in the absence of turbulence.
Indeed, turbulence disperses chemicals by advecting particles along
chaotic trajectories.
This rapidly causes large concentration gradients that speed up their
mixing down toward the smallest scales.
Turbulent mixing is a complicated and rich process; see Ref.~\cite{FGV01}
for a comprehensive review on this subject.
The mathematical treatment of the description of turbulent mixing is
closely related to that of turbulence itself, but it is in many
ways much simpler and provides therefore an ideal tool for making
conceptual progress in that field \cite{SS00}.

Here we are mainly interested in cases where it is meaningful to define a mean
concentration whose scale of variation is large compared with the scale
of the energy-carrying eddies.
In such cases it can be useful to describe the change in the mean
concentration by an effective turbulent diffusion tensor.
On smaller scales the change in the mean concentration can still be
described in such a way, but in that case the multiplication with a
turbulent diffusivity must be replaced with a convolution.
The turbulent diffusion tensor quantifies the effective exchange
of chemicals or other passive scalar quantities advected by the flow.
If there is a gradient in the mean concentration $\meanC$ of chemicals, there
will be a net mean flux $\meanFFFF=\overline{\uu c}$ of chemicals resulting
from a systematic correlation of fluctuations in the concentration $c$ and
the turbulent velocity $\uu$.
Here, overbars denote averaging.
Under isotropic conditions with sufficient scale separation,
this mean flux will be down the gradient of concentration, with
\EQ
\meanFFFF=-\kappat\nab\meanC,
\label{Fick}
\EN
where $\kappat$ is the turbulent diffusivity.
However, modifications are expected when the turbulence is anisotropic.
In that case this relation takes the form
\EQ
\meanFFF_i=-\kappa_{ij}\nabla_j\meanC,
\label{Fick2}
\EN
where $\kappa_{ij}$ is now the turbulent diffusion tensor.
In this paper we are interested in the anisotropy caused by the
presence of shear.
One of the results one expects to see is a suppression of
turbulent transport in the cross-stream direction.
This effect is discussed in various physical circumstances such as
geophysical flows \cite{Ter00}, turbulent plasmas \cite{Bur97},
and solar physics \cite{Kim05,LK06}.

Much of this research is done using analytical techniques such as the
first-order smoothing approximation and the renormalization group
analysis.
However, in recent years it has become possible to calculate turbulent
transport coefficients using numerical realizations of turbulence from
direct simulations.
Turbulent transport coefficients can then be determined by imposing
a gradient in the passive scalar concentration and measuring the
resulting concentration fluxes \cite{BKM04}.
By imposing gradients in three different directions it is possible to
assemble all components of the turbulent diffusion tensor.

In recent years such a technique has been applied to the case of
magnetic fields whose evolution is controlled not just by turbulent
magnetic diffusion, but also by non-diffusive contributions known
as the $\alpha$ effect \cite{Sch05,Sch07}.
In this way it has been possible to investigate numerically the
effects of shear and rotation in regimes that cannot be treated
analytically.
The technique is known under the name test-field method, which refers
to the fact that this approach involves the analysis of correlations
for a set of different pre-determined test fields.
In the analogous case of passive scalars, this method is now often
referred to as test-scalar method \cite{BSV09}.

Using this method, it has recently been possible to determine the
turbulent diffusion tensor in cases where the turbulence is anisotropic
owing to the presence of either rotation or an imposed magnetic field
\cite{BSV09}.
In the case of rotation the angular velocity vector $\OO$ provides a
new element for constructing an anisotropic rank-2 tensor of the form
\cite{KRP94}
\EQ
\kappa_{ij}
=\kappa_0\delta_{ij}
+\kappaO\epsilon_{ijk}\Oh_k
+\kappaOO\Oh_i\Oh_j,
\label{kappaij}
\EN
where $\hatOO=\OO/|\OO|$ is the unit vector along the rotation axis and
$\kappa_0$, $\kappaO$, and $\kappaOO$ are functions of the flow parameters.
Note that $\Omega$ is a pseudo vector while $\kappa_{ij}$ is a proper
tensor, so all three coefficients in \Eq{kappaij} are proper scalars.
In the case of a shear flow, an obvious possible ansatz is obtained by
replacing $\OO$ with the vorticity $\meanWW=\nab\times\meanUU$, which is
also a pseudovector (or axial vector), and $\meanUU$ is the mean shear flow.
However, such an ansatz would be incomplete, because it only captures
the antisymmetric part of the velocity gradient matrix $\meanU_{i,j}$,
where a comma denotes partial differentiation.
A more natural approach would therefore be to invoke both symmetric and
antisymmetric parts of the velocity gradient matrix by writing it as
$\meanU_{i,j}=\meanSSS_{ij}+\meanAAA_{ij}$, where
\EQ
\meanSSS_{ij}=\half(\meanU_{i,j}+\meanU_{j,i}),
\EN
\EQ
\meanAAA_{ij}=\half(\meanU_{i,j}-\meanU_{j,i}).
\EN
The latter can also be written as
$\meanAAA_{ij}=-\half\epsilon_{ijk}\meanW_{k}$.
A proper rank-2 tensor can then be expressed as
\EQ
\kappa_{ij}
=\kappat\delta_{ij}
+\kappa_{\sf S}\overline{\sf S}_{ij}
+\kappa_{\sf A}\overline{\sf A}_{ij}
+\kappa_{{\sf S}{\sf S}}(\meanSSSS\,\meanSSSS)_{ij}
+\kappa_{{\sf A}{\sf S}}(\meanAAAA\,\meanSSSS)_{ij},
\label{kappaijAS}
\EN
where $\kappat$, $\kappa_{\sf S}$, $\kappa_{\sf A}$,
$\kappa_{{\sf S}{\sf S}}$, and $\kappa_{{\sf A}{\sf S}}$
are proper scalars that are again functions of the flow parameters.
In the absence of helicity,
no further rank-2 tensors can be constructed from a linear shear flow.
We return to the case with helicity in \Sec{Helicity}.

An important goal of this work is to determine the coefficients
in \Eq{kappaijAS} for a linear shear flow of the form
\EQ
\meanUU=(0,Sx,0),
\label{meanUU}
\EN
where $S=\const$ is the shear rate, which is not to be confused with
the tensor $\meanSSSS$.
For a linear shear flow given by \Eq{meanUU}, the tensors
$\meanSSSS$ and $\meanAAAA$ are constants, and their
only non-vanishing components are
\EQ
\meanSSS_{xy}=\meanSSS_{yx}=-\meanAAA_{xy}=\meanAAA_{yx}=S/2.
\EN
Note also that
\EQ
\meanSSSS^2=-\meanAAAA^2=(S/2)^2\,\diag(1,1,0),
\EN
\EQ
\meanAAAA\,\meanSSSS=-\meanSSSS\,\meanAAAA=(S/2)^2\,\diag(-1,1,0).
\EN
With these preparations we can now express all nine components of
$\kappa_{ij}$ in terms of the five coefficients in \Eq{kappaijAS}
as follows:
\EQ
\kappa_{11}=\kappat+\quarter S^2
(\kappa_{{\sf S}{\sf S}}-\kappa_{{\sf A}{\sf S}}),
\EN
\EQ
\kappa_{22}=\kappat+\quarter S^2
(\kappa_{{\sf S}{\sf S}}+\kappa_{{\sf A}{\sf S}}),
\EN
\EQ
\kappa_{33}=\kappat,
\EN
\EQ
\kappa_{12}=\half S(\kappa_{\sf S}-\kappa_{\sf A}),
\EN
\EQ
\kappa_{21}=\half S(\kappa_{\sf S}+\kappa_{\sf A}),
\EN
\EQ
\kappa_{13}=\kappa_{31}=\kappa_{23}=\kappa_{32}=0.
\EN
Given that all nine components of $\kappa_{ij}$ can be determined from
simulation data using the test-scalar method, we can use the relations
above to compute the five unknown coefficients in \Eq{kappaijAS} via
\EQ
\kappa_{\sf S}S=\kappa_{21}+\kappa_{12},\quad
\kappa_{\sf A}S=\kappa_{21}-\kappa_{12},
\EN
\EQ
\kappa_{{\sf S}{\sf S}}S^2/2=\kappa_{22}+\kappa_{11}-2\kappa_{33},
\EN
\EQ
\kappa_{{\sf A}{\sf S}}S^2/2=\kappa_{22}-\kappa_{11},
\EN
\EQ
\kappat=\kappa_{33}.
\EN
Note that combinations such as $\kappa_{\sf S}S$ and
$\kappa_{{\sf S}{\sf S}}S^2/2$ have still the same dimension as
$\kappa_{ij}$, so in the following we shall quote these combinations
in that form.

In principle it is possible to construct $\kappa_{ij}$ using also
the velocity vector $\meanUU$ itself.
However, $\meanUU$ varies in $x$ and vanishes at $x=0$.
On the other hand, we expect the components of $\kappa_{ij}$ not to depend
explicitly on position, making a construction in terms of $\meanUU$ less
favorable.
Furthermore, the tensor $\meanU_i\meanU_j$, which has only one
component in the $yy$ position, can already be constructed from
$\meanSSSS^2-\meanAAAA\,\meanSSSS=\diag(0,2,0)$,
so no new information would be added.
However, this changes when we also admit helical turbulent flows, because
then there could be tensors of the form $\meanW_i\meanU_j$ and
$\meanW_j\meanU_i$ which have components in the $yz$ and $zy$ directions.
For this reason we shall also investigate helical turbulence in some cases.

A comment regarding the case of rotation without shear is here in order.
In hindsight it might have been more natural to write \Eq{kappaij}
in terms of the antisymmetric matrix $\meanAAA_{ij}=-\half\epsilon_{ijk}\Oh_k$,
i.e.\
\EQ
\kappa_{ij}
=\kappat^\Omega\delta_{ij}
+\kappa^\Omega_{\sf A}\overline{\sf A}_{ij}
+\kappa^\Omega_{{\sf A}{\sf A}}(\meanAAAA^2)_{ij},
\EN
with coefficients that are related to those in \Eq{kappaij} via
\EQ
\kappat^\Omega=\kappa_0+\kappaOO,\quad
\kappa^\Omega_{\sf A}=-2\kappaO,\quad
\kappa^\Omega_{{\sf A}{\sf A}}=4\kappaOO.
\EN
Evidently, this representation is equivalent to that of \Eq{kappaij}.

In the rest of this paper we continue with the case of a pure shear flow.
The aim is to determine the coefficients in \Eq{kappaijAS} as functions
of flow parameters such as the Peclet number and the shear parameter.

\section{Simulations}

We simulate turbulence by solving the compressible hydrodynamic equations
with an imposed random forcing term and an isothermal equation of state,
so that the pressure $p$ is related to $\rho$ via
$p=\rho\cs^2$, where $\cs$ is the isothermal sound speed.
We consider a periodic Cartesian domain of size $L^3$.
In the presence of shear the hydrodynamic equations for $\rho$
and the departure $\UU$ from the imposed shear flow $\meanUU$
take the form,
\EQ
{\DD\ln\rho\over\DD t}=-\nab\cdot\UU,
\EN
\EQ
{\DD\UU\over\DD t}=-SU_x\yyy-\cs^2\nab\ln\rho
+\ff+\FF_{\rm visc},
\label{dUdt}
\EN
where $\DD/\DD t=\partial/\partial t+(\UU+\meanUU)\cdot\nab$ is the
advective derivative with respect to the full velocity,
$\FF_{\rm visc}=\rho^{-1}\nab\cdot2\rho\nu\SSSS$ is the viscous
force, $\nu$ is the kinematic viscosity,
${\sf S}_{ij}=\half(U_{i,j}+U_{j,i})-\onethird\delta_{ij}\nab\cdot\UU$
is the traceless rate of strain tensor of the departure
from the shear flow, and $\ff$ is a random forcing
function consisting of plane transversal waves with random wave vectors
$\kk$ such that $|\kk|$ lies in a band around a given forcing wave number
$k_{\rm f}$.
The vector $\kk$ changes randomly from one timestep to the next,
so $\ff$ is $\delta$ correlated in time.
We have carried out simulations with helical
and nonhelical forcings using the modified forcing function
\begin{equation}
\ff_{\kk}=\RRRR\cdot\ff_{\kk}^{\rm(nohel)}\quad\text{with}\quad
{\sf R}_{ij}={\delta_{ij}-\ii\sigma\epsilon_{ijk}\hat{k}_k
\over\sqrt{1+\sigma^2}},
\end{equation}
where $\ff_{\kk}^{\rm(nohel)}$ is the non-helical forcing function.
In the fully helical case ($\sigma=\pm1$) we recover the forcing function used in
Ref.~\cite{B01}, and in the non-helical case ($\sigma=0$) this forcing
function becomes equivalent to that used in Ref.~\cite{HBD03}.
The forcing amplitude is chosen such that the Mach number, $\Ma=\urms/\cs$,
is about 0.1.
We use triply-periodic boundary conditions, except that the $x$ direction
is shearing--periodic, i.e.\
\EQ
\UU(-\half L,y,z,t)=\UU(\half L,y+LSt,z,t),
\EN
where $L$ is the side length of the cubic domain.
This condition is routinely used in numerical studies of shear flows
in Cartesian geometry \cite{WT88,HGB95}.

In this paper we are interested in the turbulent mixing of a passive scalar
concentration $C$.
Its evolution is governed by the equation
\EQ
{\partial C\over\partial t}=-\nab\cdot(\UU C)+\kappa\nabla^2 C,
\EN
where $\kappa$ is the microscopic (molecular) passive scalar diffusivity.
In the absence of any sources, the dynamics of $C$ depends essentially
on initial conditions.
For example, if $C$ is initially concentrated in a plane with its normal
pointing in one of the three coordinate directions, turbulence tends to
spread this initial distribution away from the plane -- regardless of
its orientation.
Only the speed of spreading will be different in the different directions.
The spreading is then best described by introducing planar averages
over the same directions as the initial distribution.
These averages are denoted by overbars and they
depend only on time and the direction normal to the plane of averaging,
i.e.\ $\meanC=\meanC(x_j,t)$, where $x_j$ denotes $x$, $y$, or $z$ for
$j=1$, ..., 3, just depending on the initial distribution.
This allows us then to quantity the speed of spreading by the different
components of the diffusion tensor $\kappa_{ij}$ in \Eq{Fick2}.
We do this by introducing different `test scalars' and calculating
the evolution for each case separately..

\begin{figure*}[t!]\begin{center}
\includegraphics[width=\textwidth]{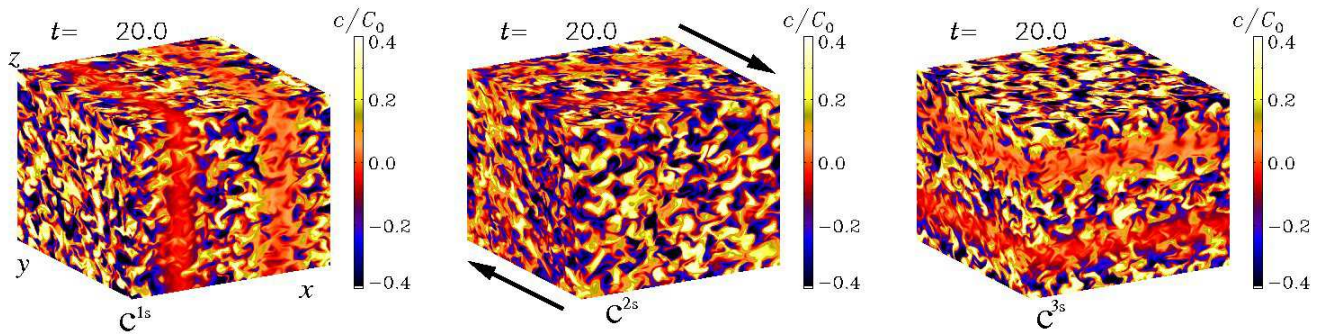}
\end{center}\caption[]{
(Color online)
Visualization of $c^{1s}$, $c^{2s}$, and $c^{3s}$ on the periphery
of the computational domain after about one turnover time for a
run with $k/\kf=0.1$.
In the middle panel, arrows indicate the direction of the shear flow
with negative $S$, i.e.\ $\dd\meanU_y/\dd x<0$.
Note the clear sinusoidal modulation in the $x$, $y$, and $z$
directions for the three panels, respectively.
In the middle panel this modulation is already smeared out by the shear.
}\label{c}\end{figure*}

In the following we are interested in the fluxes of the passive scalar
concentration, $\meanFFFF=\overline{\uu c}$, where $c=C-\meanC$ is the
fluctuation around the mean concentration and $\uu=\UU-\meanUU$ is the
velocity fluctuation around the mean flow $\meanUU$.
The test-scalar equation is obtained by subtracting the averaged passive
scalar equation from the original one and applying it to a predetermined
set of six different mean fields,
\EQ
\meanC^{ic}=C_0\cos kx_i,\quad\meanC^{is}=C_0\sin kx_i,\\
\label{TestScalar}
\EN
where $C_0$ is a normalization factor.
Again, the overbars denote planar averaging over the directions that are
perpendicular to the direction in which the mean field varies.
For each test field $\meanC^{pq}$ we obtain a separate evolution
equation for the corresponding fluctuating component $c^{pq}$,
\EQ
{\partial c^{pq}\over\partial t}=
-\nab\cdot(\meanUU c^{pq}+\uu\meanC^{pq}+\uu c^{pq}-\overline{\uu c^{pq}})
+\kappa\nabla^2c^{pq},
\label{dcpqdt}
\EN
where $p=1$, ..., 3, and $q=c$ or $s$.
In this way, we calculate six different fluxes,
$\meanFFFF^{pq}=\overline{\uu c^{pq}}$, and compute
the nine relevant components of $\kappa_{ij}$,
\EQ
\kappa_{ij}=-\bra{\cos kx_j\meanFFF_i^{js}-\sin kx_j\meanFFF_i^{jc}}/k,
\label{Fxyz}
\EN
for $i,j=1, ..., 3$.
Here, angular brackets denote volume averages.
A visualization of $c^{1s}$, $c^{2s}$, and $c^{3s}$ on the periphery of
the computational domain is shown in \Fig{c} after about one turnover
time for a run with $k/\kf=0.1$, which is smaller than in most of the
runs analyzed in this paper.
This ratio is chosen here for visualization purposes only, because
this way the large-scale modulation compared with the scale of the
turbulence becomes evident.

We emphasize that \Eq{dcpqdt} is an inhomogeneous equation in $c^{pq}$.
The term $\uu\meanC^{pq}$ can be regarded as a forcing term that
guarantees that the direction of the turbulent concentration flux will
not change with time.

In this paper we present the values of $\kappa_{ij}$ in non-dimensional
form by normalizing with
\EQ
\kappatz=\urms/3\kf,
\label{kappatz}
\EN
which is the expected value for large values of $\Pe$.
Here we have defined the root-mean-square value of the velocity
fluctuation as $\urms=\bra{\uu^2}^{1/2}$.

Our simulations are characterized by two important non-dimensional
control parameters, the shear parameter $\Sh$ and the Peclet number
$\Pe$, defined as
\EQ
\Sh=S/(\urms\kf),\quad
\Pe=\urms/(\kappa\kf).
\EN
In addition, there is the Schmidt number $\Sc=\nu/\kappa$, but we keep
it equal to unity in all cases reported below.
Note also that in most cases we use negative values of $S$,
so we have $\Sh<0$.
The smallest wave number that fits into the computational domain
is $k_1=2\pi/L$.
In most of the cases reported below
we choose the forcing wave number to be 3 times larger, i.e.\ $\kf/k_1=3$.

The simulations have been carried out using the {\sc Pencil
Code} \footnote{http://pencil-code.googlecode.com/} which
is a high-order finite-difference code (sixth order in space and third
order in time) for solving the compressible hydrodynamic equations.
The test-scalar equations where already implemented into the public-domain
code, but have now been generalized to determining all nine components of
$\kappa_{ij}$.
The numerical resolution used in the simulations depends on the Peclet
number and reaches $128^3$ meshpoints for runs with $\Pe\approx120$.
In this paper we restrict ourselves to time spans short enough so
that the so-called vorticity dynamo has no time do develop; see
Refs.~\cite{EKR03,KMB09} for details on this effect.

\section{Results}

\subsection{Dependence on the shear parameter}

\begin{figure}[t!]\begin{center}
\includegraphics[width=\columnwidth]{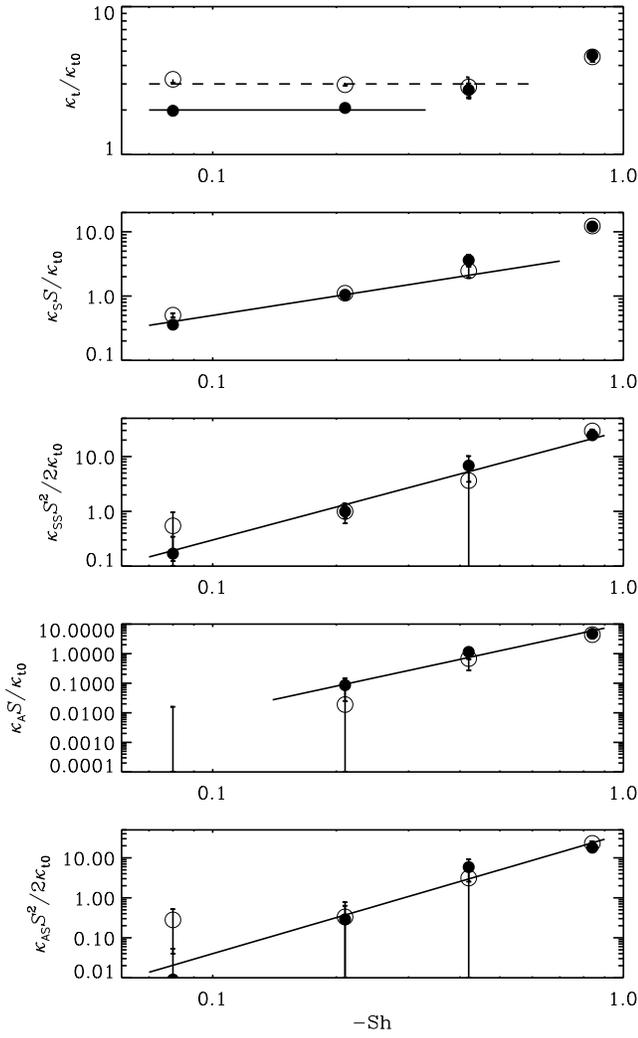}
\end{center}\caption[]{
Dependence of the coefficients in \Eq{kappaijAS} on Sh for $\Pe=25$.
The dashed line in the first panel is for a run with maximum helicity.
All runs with helicity are marked with open symbols.
Filled symbols indicate runs without helicity.
Solid lines represent the fits given by \Eqs{fits1a}{fits2a}.
}\label{pkappaAS1_Sh}\end{figure}

We begin by discussing the dependence of the coefficients in
\Eq{kappaijAS} on the shear parameter $\Sh$.
The result is shown in \Fig{pkappaAS1_Sh} for $\Pe=25$.
It turns out that all five coefficients are positive.
We find that $\kappat/\kappatz=\const=2$ for non-helical turbulence and 3 for
helical turbulence, independent of the value of shear, provided $|\Sh|<0.5$.
The other coefficients show the following approximate scaling behavior:
\EQ
\kappa_{\sf S}S/\kappatz\approx5|\Sh|,\quad
\kappa_{{\sf S}{\sf S}}S^2\!/2\kappatz\approx30\,\Sh^2,
\label{fits1a}
\EN
\EQ
\kappa_{\sf A}S/\kappatz\approx10\,|\Sh|^3,\quad
\kappa_{{\sf A}{\sf S}}S^2/2\kappatz\approx40\,|\Sh|^3.
\label{fits2a}
\EN
The fact that $\kappa_{\sf A}$ and $\kappa_{{\sf A}{\sf S}}$
scale with the third power of $\Sh$ suggests that these are
higher order effects that are not easily captured by perturbative
approaches.

A comment regarding the values of $\Sh$ is here in order.
Although values of $\Sh$ larger than unity have not yet been
explored, it is unlikely that the uprise of $\kappat$ continues.
Furthermore, one might speculate that all coefficients in \Eqs{fits1a}{fits2a}
should eventually decrease as $|\Sh|\to\infty$.

In \Fig{pkappaAS1_Sh} we have also shown results for cases where the forcing
function has maximum helicity.
No significant dependence can be seen, except for $\kappat$ which is
slightly enhanced in the helical case with weak shear.
This suggests that this dependence is not connected with the presence
of shear.

\subsection{Dependence on Peclet number}

\begin{figure}[t!]\begin{center}
\includegraphics[width=\columnwidth]{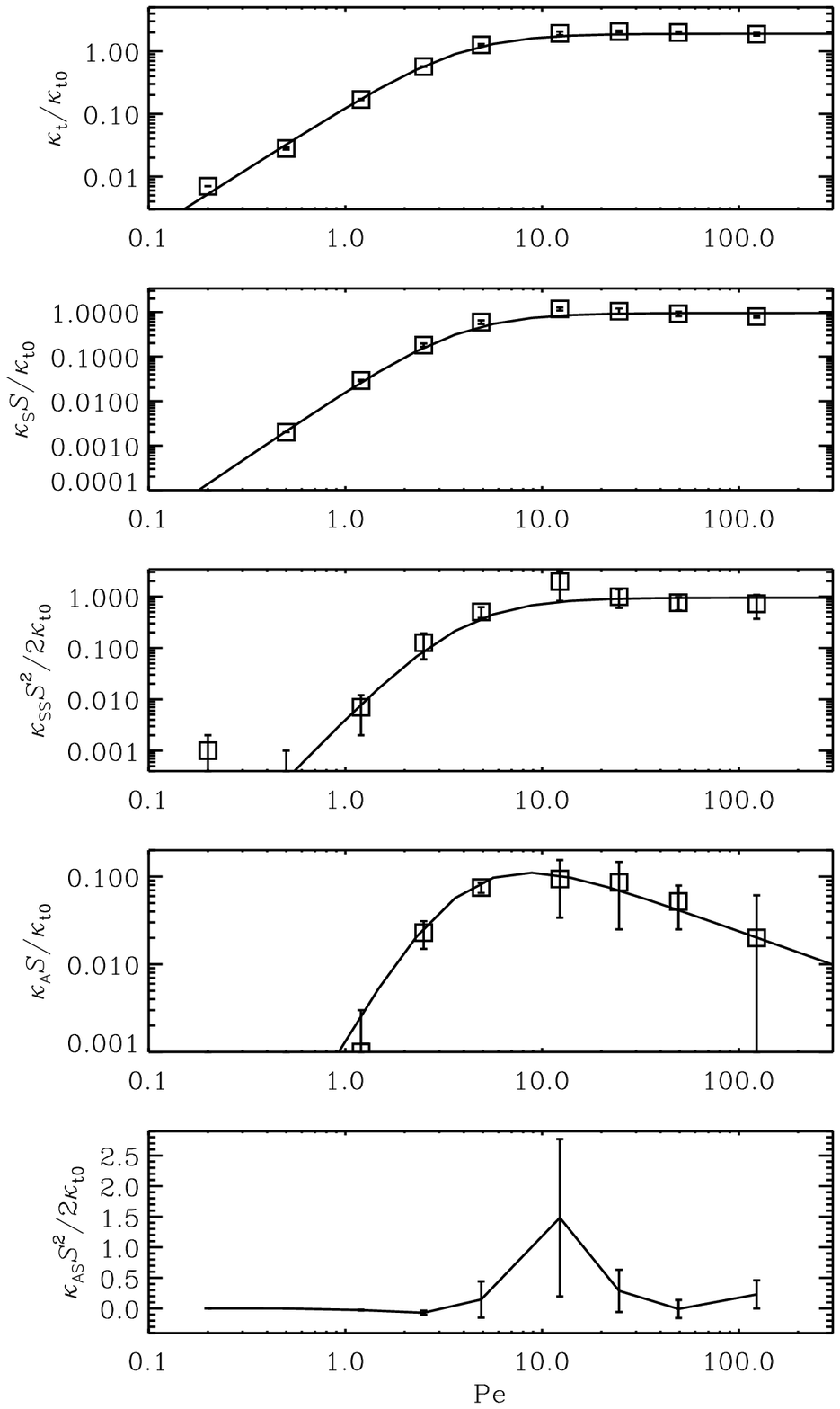}
\end{center}\caption[]{
Dependence of the coefficients in \Eq{kappaijAS} on Pe for $\Sh=-0.2$.
The symbols give the numerical results and the solid lines represent fits
given by \Eqs{fits1}{fits2}.
}\label{pkappaAS1}\end{figure}

We have performed simulations for different values of the Peclet number
and have determined the coefficients in \Eq{kappaijAS} for each simulation.
The results are shown in \Fig{pkappaAS1} for fixed $\Sh=0.2$.
It turns out that the first four coefficients can well be approximated
by simple algebraic functions,
\EQ
{\kappat\over\kappatz}={2\kappa_{\rm Sh}\Pe^2\over\Pe_0^2+\Pe^2},\quad
{\kappa_{\sf S}S\over\kappatz}={\kappa_{\rm Sh}\Pe^3\over(\Pe_0^2+\Pe^2)^{3/2}},
\label{fits1}
\EN
\EQ
{\kappa_{{\sf S}{\sf S}}S^2\over2\kappatz}
={\kappa_{\rm Sh}\Pe^4\over(\Pe_0^2+\Pe^2)^2},\quad
{\kappa_{\sf A}S\over\kappatz}
={\kappa_{\rm Sh}\Pe^4\over(\Pe_0^2+\Pe^2)^{2.4}},
\label{fits2}
\EN
where $\kappa_{\rm Sh}=0.95\kappa_{\rm t0}$ and $\Pe_0=3.8$ are fit parameters.
In the case of $\kappa_{{\sf A}{\sf S}}$ the error bars are so large that
no conclusive statements can be made.
Likewise, the error bar on the first data point is quite large too.
This is caused by the numerical time step becoming rather short at large
diffusivities, so the run is short and the statistics poor.

\begin{figure}[t!]\begin{center}
\includegraphics[width=\columnwidth]{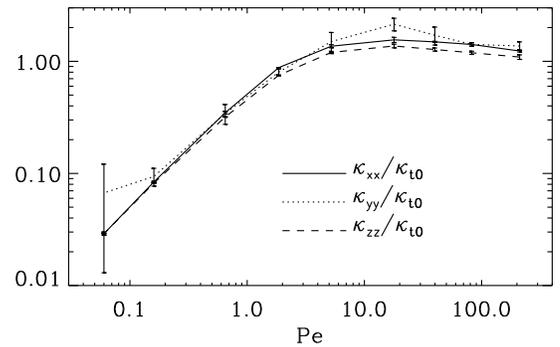}
\end{center}\caption[]{
Dependence of the diagonal components of $\kappa_{ij}$ on $\Pe$.
}\label{pkappa1}\end{figure}

In \Fig{pkappa1} we show the dependence of the diagonal components
of $\kappa_{ij}$ on $\Pe$.
Over the range of parameters shown here, the difference
between the three components is small, although there is a
tendency for $\kappa_{yy}$ to be somewhat enhanced around $\Pe=20$,
while $\kappa_{zz}$ is slightly smaller than $\kappa_{xx}$.

\subsection{Wavenumber dependence}

We consider now the dependence of the diagonal components of $\kappa_{ij}$
on the wave number $k$ of the test scalar in \Eq{TestScalar}.
A dependence of $\kappa_{ij}$ on $k$ reflects the fact that there is
poor scale separation, i.e.\ $k/\kf$ is no longer small.
In such a case, the multiplication with a turbulent diffusivity
in \Eqs{Fick}{Fick2} must be replaced by a
convolution with an integral kernel \cite{BSV09}.
In Fourier space the convolution corresponds to a multiplication.
The full integral kernel can be assembled by determining the full
$k$ dependence and then Fourier transforming back into real space.

\begin{figure}[t!]\begin{center}
\includegraphics[width=\columnwidth]{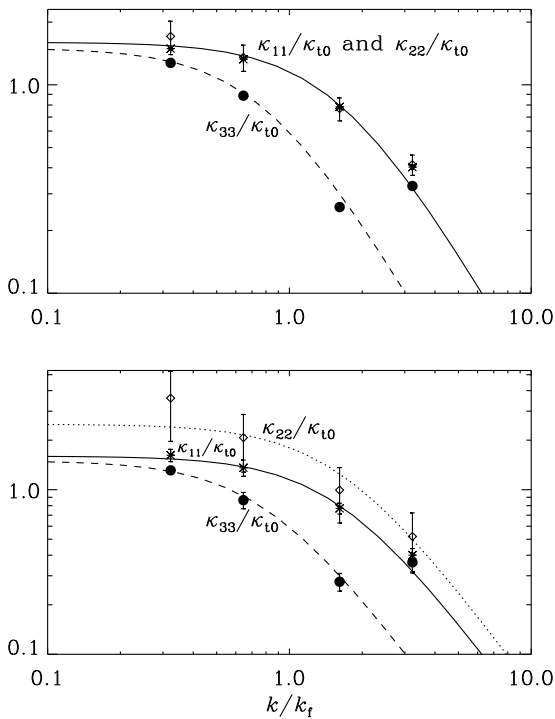}
\end{center}\caption[]{
Dependence of the diagonal components of $\kappa_{ij}$ on $k$
for $\Sh=-0.13$ at $\Pe=40$ (upper panel) and $\Sh=-0.20$ at
$\Pe=60$ (lower panel).
}\label{pkappa1_k}\end{figure}

The resulting dependence on $k$ is shown in \Fig{pkappa1_k} for
two values of the shear parameter and $\Pe$ around 50.
In agreement with earlier findings, the components of $\kappa_{ij}$
show a Lorentzian dependence on $k$, i.e.\
\EQ
\kappa_{ij}={\kappa_{ij}^{(0)}\over1+(a k/\kf)^2},
\label{Lorentzian}
\EN
where $a\approx0.2$ for the $\kappa_{11}$ and $\kappa_{22}$ components,
and $a\approx0.4$ for the $\kappa_{33}$ component.
Here, $\kappa_{ij}^{(0)}$ is the value for $k=0$, which is approximately
equal to $\kappatz$, defined in \Eq{kappatz}.

Given that the Schmidt number is always kept equal to unity, there
will be a fully developed cascade in the passive scalar concentration
when the Peclet number is large.
The validity of \Eq{Lorentzian} has only been tested for values of
$\Pe$ up to 60.
It is unclear whether this equation holds also for large values of $\Pe$
when contributions from the high wave number dynamics may become
important in the mixing of the mean concentration.

The case of high wave numbers is interesting in view of possible
applications of our results to subgrid scale modeling in large-eddy
simulations of turbulence.
The highest possible wave number is the Nyquist wave number,
$k_{\rm Ny}=\pi/\delta x$, where $\delta x$ is the mesh scale.
In the Smagorinsky model \cite{Sma63} the subgrid scale viscosity is
proportional to the modulus of the rate of strain tensor times $\delta x^2$.
For a turbulent flow where the local velocity difference $\delta u_\ell$
over a distance $\ell$ is proportional to $\ell^{1/3}$ we expect the
subgrid scale viscosity to be effectively proportional to $\ell^{4/3}$,
suggesting an asymptotic $k^{-4/3}$ scaling for $k\gg\kf$.
Here we have identified $\ell$ with $\delta x$ and thus $k$ with $k_{\rm Ny}$.
Only for a smooth velocity field, where $\delta u_\ell$ scales linearly
with the separation $\ell$, the subgrid scale viscosity would be
proportional to $\ell^2$, justifying an asymptotic $k^{-2}$ scaling.
This uncertainty warrants further studies of the validity of
\Eq{Lorentzian} for $k\gg\kf$.

\subsection{Effects of helicity}
\label{Helicity}

\begin{figure}[t!]\begin{center}
\includegraphics[width=\columnwidth]{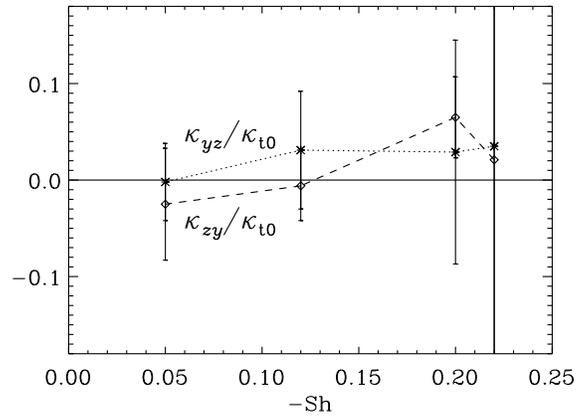}
\end{center}\caption[]{
Plot of $\kappa_{yz}$ (dotted line) and $\kappa_{zy}$ (dashed line) versus $\Sh$
for maximally helical turbulence and $\Pe=25$.
No significant dependence can be seen.
}\label{pkappa1_HSh}\end{figure}

As discussed in the Introduction, the presence of helicity allows one
in principle to construct proper tensors proportional to
$\meanW_i\meanU_j$ and $\meanW_j\meanU_i$, because we have now
access to a pseudoscalar given by the kinetic helicity of the turbulence.
If this does indeed have an effect, one would expect
finite $yz$ and $zy$ components.
In \Fig{pkappa1_HSh} we present results for $\kappa_{yz}$ and
$\kappa_{zy}$ using $\Pe=25$.
We see that $\kappa_{yz}=\kappa_{zy}=0$ within error bars,
so there is no evidence for the presence of additional
terms when the turbulence is helical.

\section{Expectations from the $\tau$ approximation}
\label{KappaA}

Passive scalar transport is closely related to the transport of a
mean magnetic field.
Commonly applied techniques for computing turbulent transport coefficients
in mean-field electrodynamics are the first order smoothing approximation
\cite{Mof78,KR80} and the $\tau$ approximation \cite{VK83,KMR96,RKR03}.
The $\tau$ approximation consists in writing down an evolution equation
for the quadratic correlations which, in the case of mean-field
electrodynamics, is the mean electromotive force $\meanEMF$.
Its solution gives then an expression for $\meanEMF$ in terms of the
mean magnetic field and its derivatives.
For a recent review see Ref.~\cite{BS05}.
This technique has also been used to compute the Reynolds and Maxwell
stress in rotating shear flows \cite{Ogi03,GO05,LiljeO09}.
In the present case of passive scalar transport one starts with
the evolution equation for the mean flux $\meanFFFF=\overline{\uu c}$,
as is done in Refs.~\cite{BF03,BKM04}.
Thus, we write
\EQ
{\partial\meanFFF_i\over\partial t}
=\overline{\dot{u}_i c}+\overline{u_i\dot{c}},
\EN
where dots denote time derivatives that are given essentially by
\Eqs{dUdt}{dcpqdt}.
This results in quadratic and triple correlations.
The sum of all triple correlations is substituted by a damping term
of the form $-\meanFFFF/\tau$ on the right-hand side of the evolution
equation for $\meanFFFF$.
Here, $\tau=\St/\urms\kf$ is the turnover time and $\St$ is a positive
dimensionless parameter of order unity (referred to as Strouhal number).
This is a closure assumption that cannot be motivated rigorously \cite{RR07},
but it has been found numerically that the triple-correlations are indeed
locally and temporally proportional to the negative flux term divided by
$\tau$; see Ref.~\cite{BKM04} for passive scalar diffusion and
Ref.~\cite{BS05b} for the case of mean-field electrodynamics.

\begin{figure}[t!]\begin{center}
\includegraphics[width=\columnwidth]{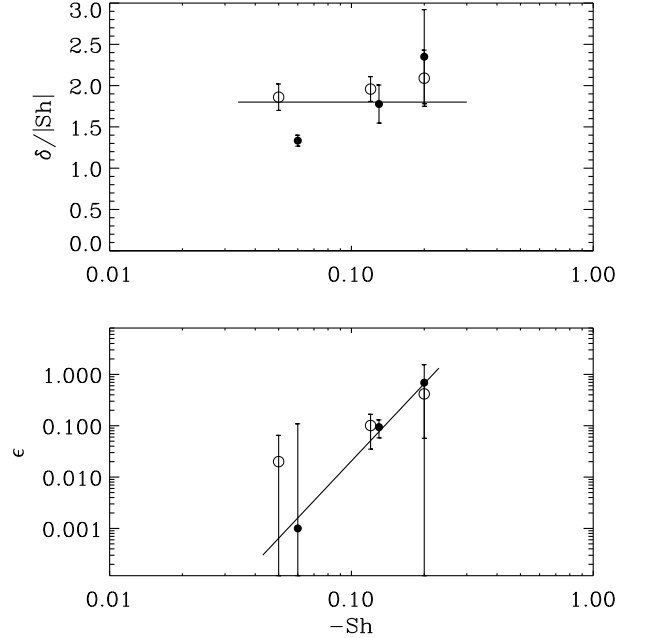}
\end{center}\caption[]{
Dependence of $\delta$ and $\epsilon$ on $\Sh$ for nonhelical turbulence
(solid symbols) and helical turbulence (open symbols).
The solid line in the second panel has a slope of 5.
}\label{pkappa1_uu}\end{figure}

As a first orientation, and in order to gain some understanding of
our numerical results, we make the additional assumption that we can
subsume the effects of the pressure term in our closure assumption.
Since our forcing function $\ff$ is $\delta$ correlated in time
we have $\overline{\ff c}=\bm{0}$ and thus obtain
\EQ
\overline{\dot{u}_i c}=-S\delta_{i2}\delta_{1k}\overline{u_kc}
+\mbox{triple correlations},
\EN
\EQ
\overline{u_i \dot{c}}=-\overline{u_iu_j}\,\nabla_j\meanC
+\mbox{triple correlations}.
\EN
The triple correlation terms result from the nonlinearities in the
evolution equations, \Eqs{dUdt}{dcpqdt}.
In the $\tau$ approximation one substitutes the sum of the triple
correlations by quadratic correlations, i.e.\ in the present case 
by $-\overline{u_ic}/\tau$ \cite{VK83,KRR90}.
We write the resulting equation in matrix form,
\EQ
\tau{\partial\meanFFF_i\over\partial t}=-{\sf L}_{ik}\meanF_k
-\tau\overline{u_iu_j}\,\nabla_j\meanC,
\EN
where ${\sf L}_{ik}=\delta_{ik}+S\tau\delta_{i2}\delta_{1k}$.
We solve this equation for $\meanFFFF$ and obtain
\EQ
\meanF_i=-(\LLLL^{-1})_{ij}\left(\tau\overline{u_ju_k}\,
\nabla_k\meanC+\tau{\partial\meanF_i\over\partial t}\right),
\label{meanF2}
\EN
where $(\LLLL^{-1})_{ik}=\delta_{ik}-\Sh\,\delta_{i2}\delta_{1k}$
with $\Sh=S\tau$.
In the presence of shear, the Reynolds stress tensor $\overline{u_ju_k}$
is no longer diagonal, but it has finite $xy$ and $yx$ components.
Also the three diagonal components are no longer the same.
In the following we represent $\overline{u_ju_k}$ in the form
\EQ
\overline{\uu\uu}=\overline{u_x^2}
\pmatrix{1 & -\delta & 0 \cr -\delta & 1+\epsilon & 0 \cr 0 & 0 & 1+\epsilon_z},
\EN
where $\delta=-\overline{u_xu_y}/\overline{u_x^2}$ characterizes the
value of the off-diagonal components, while
$\epsilon=\overline{u_y^2}/\overline{u_x^2}-1$ and
$\epsilon_z=\overline{u_z^2}/\overline{u_x^2}-1$ characterize
the change in the two lower diagonal components.
The dependence of $\delta$ and $\epsilon$ on $\Sh$ is shown
in \Fig{pkappa1_uu}, while $\epsilon_z$ is found to be small.
Inserting this expression into \Eq{meanF2}, we obtain
\EQ
{{\bm\kappa}\over\kappatz}=\pmatrix{
1 & -\delta & 0 \cr
-\delta-\Sh & 1+\epsilon+\delta\Sh & 0 \cr
0 & 0 & 1+\epsilon_z}.
\EN
In the stationary state we may ignore the time derivative
and recover \Eq{kappaij} with
\EQ
{\kappa_{\sf S}S\over\kappatz}=-2\delta-\Sh,\quad
{\kappa_{\sf A}S\over\kappatz}=-\Sh,
\label{kappaOO}
\EN
\EQ
{\kappa_{\sf SS}S^2\!/2\over\kappatz}+2\epsilon_z=
{\kappa_{\sf AS}S^2\!/2\over\kappatz}=\epsilon-\delta\Sh.
\EN
We recall that $\Sh$ is negative, and that $\delta$ changes sign with $\Sh$.
Therefore we expect $\kappa_{\sf S}$ and $\kappa_{\sf A}$ to be positive,
which agrees with the simulations.
Furthermore, we expect $\kappa_{yy}$ to be enhanced, which also agrees
with the simulations.
However, the slight suppression of $\kappa_{zz}$ cannot be explained by
the simple theory, because $\epsilon_z$ is small and perhaps even positive,
suggesting at best an opposite trend.

\section{Conclusions}

The present work has shown that shear introduces anisotropies
in the diffusivity tensor for passive scalar diffusion.
These additional components are proportional to the even and
odd parts of the velocity gradient tensor, as well as products
of these tensors.
Those components that are connected with the antisymmetric part
of the velocity gradient tensor scale with the third power of the
shear parameter, suggesting that these effects cannot be captured
perturbatively.

Given that $\Sc=1$ in all our runs, we always have $\Rey=\Pe$, which is
at most about 100, so the inertial range of the turbulence is not very
big yet.
It is therefore important to investigate the dependence of the various
transport coefficients on the values of $\Rey$ and $\Pe$, as was done
in \Fig{pkappaAS1}.
The results available so far suggest that the first three coefficients
($\kappat$, $\kappa_{\sf S}$, and $\kappa_{{\sf S}{\sf S}}$) do not change
with $\Rey$ for $\Rey>10$.
If there were indications that the resulting coefficients change beyond
$\Rey=100$, it would be important to make an effort to increase the values
of $\Rey$ even further.
This would require more resolution and is obviously expensive.
In view of the constancy of the first three coefficients, this may not
be well justified.
The fourth coefficient ($\kappa_{\sf A}$) seems to tend to zero, and
the fifth one ($\kappa_{{\sf A}{\sf S}}$) shows large error bars.
The situation regarding these last two coefficients may not improve
significantly towards larger Reynolds numbers, unless the simulations
are run for long enough time.

In general, turbulent transport tends to be enhanced in the direction
of the shear, i.e.\ $\kappa_{yy}$ tends to be larger than
$\kappa_{xx}$ and $\kappa_{zz}$.
Furthermore, $\kappa_{zz}$ tends to be suppressed relative to
$\kappa_{xx}$.
This is a result that is not reproduced by a simple analytical closure
in which triple correlations are being replaced with quadratic ones.
In particular, there is no evidence for a suppression of turbulent
transport in the cross-stream or $x$ direction.
Instead, there is a suppression in the spanwise direction out of the plane
of the shear flow.

We recall that the moduli of the diagonal components of the turbulent
diffusivity tensor are found to decrease with increasing wave number of
the mean concentration in a Lorentzian fashion.
This is in agreement with earlier findings both in the contexts of
mean-field electrodynamics with and without shear \cite{BRS08,Mitra_etal09},
as well as passive scalar transport in the absence of shear \cite{BSV09}.
The limit of high wave numbers may be of interest for subgrid scale modeling
in large-eddy simulations of turbulence.
However, it still needs to be clarified whether the effective diffusivity
is proportional to the inverse Nyquist wave number to the second power,
as suggested by our current results, or to some smaller power,
$\sim k^{-4/3}$, as expected for Kolmogorov turbulence.
In order to address this question, simulations at larger
Peclet and Reynolds numbers are required.
Such simulations do not require the presence of shear.
This is however beyond the scope of the present paper.

Finally, we note that, in shear flows, the passive scalar transport
properties are not affected by the presence of helicity.
In other words, there is no evidence for the existence of components
to the turbulent diffusivity tensor $\kappa_{ij}$ that are proportional
to $\meanW_i\meanU_j$ and $\meanW_j\meanU_i$.

\acknowledgments
We thank Alexander Hubbard and Karl-Heinz R\"adler for suggestions and
stimulating discussions.
We acknowledge the use of computing time at the Center for
Parallel Computers at the Royal Institute of Technology in Sweden.
This work was supported in part by
the European Research Council under the AstroDyn Research Project No.\ 227952
and the Swedish Research Council Grant No.\ 621-2007-4064.

%r e f

\vfill\bigskip\noindent\tiny\begin{verbatim}
$Header: /var/cvs/brandenb/tex/eniko/testscalar_shear/paper.tex,v 1.32 2010-07-02 19:11:29 brandenb Exp $
\end{verbatim}

\end{document}